\documentclass[twocolumn,showpacs,unsortedaddress,superscriptaddress,prc,aps,10pt,floatfix]{revtex4-1}
\usepackage{pstricks}
\usepackage{graphicx}
\usepackage{bm}
\usepackage{subfig}
\usepackage{amsmath}
\usepackage{mathrsfs}
\newcommand{\anl}{Argonne National Laboratory, Argonne, IL 60439, USA}
\newcommand{\cenpa}{Center for Experimental Nuclear Physics and Astrophysics,
University of Washington, Seattle, WA 98195, USA}
\newcommand{\lpc}{Normandie Univ, ENSICAEN, UNICAEN, CNRS/IN2P3, LPC Caen, 14000 Caen, France}
\newcommand{\nscl}{National Superconducting Cyclotron Laboratory and
Department of Physics and Astronomy,
Michigan State University, East Lansing, MI 48824, USA}
\begin{document}
\title{Charge-state distribution of Li ions from the $\beta$ decay of laser-trapped $^{6}$He atoms}


\author{R.~Hong}
\altaffiliation[Corresponding author: ]{hongran@uw.edu}
\affiliation{\cenpa}\affiliation{\anl}

\author{A.~Leredde}
\affiliation{\anl}

\author{Y.~Bagdasarova}
\affiliation{\cenpa}

\author{X.~Fl\'echard}
\affiliation{\lpc}

\author{A.~Garc\'{i}a}
\affiliation{\cenpa}

\author{A.~Knecht}
\altaffiliation[Present address: ]{Paul Scherrer Institute,
Villigen PSI, 5232, Switzerland}
\affiliation{\cenpa}
 
\author{P.~M\"uller}
\affiliation{\anl}

\author{O.~Naviliat-Cuncic}
\affiliation{\nscl}

\author{J.~Pedersen}
\affiliation{\cenpa}

\author{E.~Smith}
\affiliation{\cenpa}

\author{M.~Sternberg}
\altaffiliation[Present address: ]{Porch,
2200 1st Ave South
Seattle, WA 98134}
\affiliation{\cenpa}

\author{D.\,W.~Storm}
\affiliation{\cenpa}

\author{H.\,E.~Swanson}
\affiliation{\cenpa}

\author{F.~Wauters}
\altaffiliation[Present address: ]{Institut f\"ur Kernphysik, Johann-Joachim-Becherweg 45,
Johannes Gutenberg-Universit\"at Mainz,
55128 Mainz, Germany}
\affiliation{\cenpa}

\author{D.~Zumwalt}
\altaffiliation[Present address: ]{Porch,
2200 1st Ave South
Seattle, WA 98134}
\affiliation{\cenpa}

\date{\today}
\begin{abstract}
The accurate determination of atomic final states following nuclear $\beta$ decay plays an important role in several experiments. In particular, the charge state distributions of ions following nuclear $\beta$ decay are important for determinations of the $\beta-\nu$ angular correlation with improved precision. Beyond the hydrogenic cases, the decay of neutral $^6{\rm He}$ presents the simplest case. Our measurement aims at providing benchmarks to test theoretical calculations. The kinematics of Li$^{n+}$ ions produced following the $\beta$ decay of $^6{\rm He}$ within an electric field were measured using $^6{\rm He}$ atoms in the metastable $(1s2s,~{^3S_1})$ and in the $(1s2p,~{^3P_2})$ states confined by a magneto-optical trap. The electron shake-off probabilities were deduced including their dependence on ion energy. We find significant discrepancies on the fractions of Li ions in the different charge states with respect to a recent calculation.\\

\end{abstract}


\maketitle
\section{Introduction}
\label{Sec_Intro}

Atomic and molecular degrees of freedom can play an important role in precision nuclear beta-decay experiments. In nuclear beta decays, two energetic leptons (an electron and an anti-neutrino) are emitted, while the daughter nucleus recoils. The nucleus is usually in an atom or molecule, and the sudden change of its charge and its recoiling motion may cause electron excitations, shake-offs and molecular excitations. The final state of the recoil ion affects the shape of the $\beta$-energy spectrum. For example, in measurements of the $\beta$-energy spectrum near the end-point from molecular tritium, the final electronic state distribution \cite{Jon99,Sae00,Dos06,Bod15} can affect the determination of the mass of anti-neutrinos. 
The helicity properties of the weak interaction imply correlations between the momenta of the outgoing particles \cite{rh_helicity}. Thus, precise measurements of the $\beta-\nu$ angular correlation can be used to search for new interactions\cite{Nab,Sci04,beh09,go:05}. In such experiments the $\beta-\nu$ angular correlation coefficient is deduced from the kinematics of the recoil ion, which can depend on the molecular binding\cite{Sci04,beh09,Vor03}. 

The beta decay of $^6{\rm He}$ presents a good opportunity to determine the $\beta-\nu$ correlation. Because of the large endpoint and the relatively light mass of the nucleus, the $\beta-\nu$ correlation has a significant effect on the kinematics of the charged particles from the decay. A measurement performed in 1963 \cite{Joh63} was one of several landmark experiments that determined the $V-A$ nature of the weak interaction: the charged weak currents are of vector and axial vector type. Fundamental measurements of this kind have renewed interest in the context of searching for hints of new physics as deviations from the expectations based on the Standard Model \cite{ci:13}. An ongoing experiment is aiming at a measurement of the $\beta-\nu$ angular correlation in $^6{\rm He}$ decay with improved precision  \cite{Kne13,Leredde15}. In this experiment, the momentum of the recoil ion emitted from a cold and dilute cloud of laser-cooled $^6{\rm He}$ atoms confined in a magneto-optical trap (MOT)  \cite{Leredde15} is determined through a full kinematics reconstruction in a strong electric field. Due to the sudden change in nuclear charge, the electrons do not always find the corresponding orbits in the Li atom and can be {\em shaken off}. Thus, the $^6{\rm Li}$ ion can have electric charges between $+1$ and $+3$. The fraction in a given charge state not only depends on the overlap of the initial electronic wave function and the final continuum states, but also on the ion energy, so a proper extraction of the $\beta-\nu$ correlation coefficient requires understanding the shake-off effect. 

Quantitative comparisons of charge distributions have been presented for two heavier systems, $^{35}$Ar \cite{Couratin13} and $^{21}$Na \cite{Scielzo03}. The calculations in  $^{21}$Na did not take into account a potentially important cancellation factor in the recoil-energy-dependent fraction that we describe below. The comparisons in $^{35}$Ar showed agreement at the 1\% level. As we indicate below the $^{6}$He system invites for higher precision comparisons, because the electronic wave functions for helium can be calculated with high accuracy. The case of $^6{\rm He}$ also presents a nice benchmark to test aspects of the calculations that are relevant for other problems, like the role of electron-electron interactions, the use of the sudden approximation, and methods for calculating charge distributions after the shake-off process. Several calculations have been performed for $^6{\rm He}$ \cite{Wauters96,Frolov10,Sch15}. 
A confirmation of the calculated fraction of ${\rm Li}^{3+}$ was recently performed with the hydrogen-like system of $^6{\rm He}$ ions \cite{Cou12}. Here, we address the case of the $^6{\rm He}$ neutral atom, which presents additional ingredients associated with the two electrons. 

\begin{table}[!h]
\caption{Comparison of calculated versus measured $^6{\rm Li}$ ion charge fractions (in \%) for $^6{\rm He}$ decays from the $^{1}S$ atomic ground state.}
\begin{ruledtabular}
\begin{tabular}{lll}
Ion & \multicolumn{1}{c}{Theory\footnotemark[1] \cite{Sch15}} & \multicolumn{1}{l}{Previous}\\
& \multicolumn{1}{c}{}& \multicolumn{1}{l}{Experiment \cite{Car63}}\\
\hline
${\rm Li}^+$  & $88.99(2)$  & $89.6(2)$  \\
${\rm Li}^{2+}$  & $9.7(1)$ & $10.4(2)$ \\
${\rm Li}^{3+}$  & $1.2(1)$ & $0.042(7)$\\
\end{tabular}
\footnotetext[1]{In Ref.~\cite{Sch15} the energy-dependent shake-off probabilities are modeled as $P=A+B\times E_{ion}$ and the values of parameters $A$ and $B$ are presented for all charge states. We calculated the average ion energy $\langle E_{ion}\rangle$, and used it to calculate the theoretical energy-integrated charge state fraction. When calculating $\langle E_{ion}\rangle$, the $\beta-\nu$ correlation coefficient is assumed to be $-1/3$, and no $\beta$-energy threshold or directional restrictions on the emitted leptons are applied. In this case, $\langle E_{ion}\rangle=0.723$~keV. }
\label{tab:total fractions}
\end{ruledtabular}
\end{table}
The charge distribution of Li ions from the decay of $^6{\rm He}$ in its electronic ground state was measured by Carlson et al. \cite{Car63}. Table~\ref{tab:total fractions} shows a comparison to the most recent calculation of Schulhoff and Drake \cite{Sch15}. 
As can be observed, there are significant discrepancies for Li-ion fractions in the different charge states. The main aim of the calculations in Ref.~\cite{Sch15} was to study the dependence of charge distribution on the Li-ion energy, but a clear prediction is also given for the overall fractions in different charge states from $^6{\rm He}$, both in its ground and metastable states.   

We report here the first measurement of the electron emission probabilities following the $\beta$ decay of $^6{\rm He}$ atoms confined via a magneto-optical trap working between the metastable $(1s2s,~{^3S_1})$ and the $(1s2p,~{^3P_2})$ states. We present data with both the trapping lasers off (pure $^3S_1$) and on (approximate 50\%/50\% mixture of $^3S_1$ and $^3P_2$) and compare with the calculations of Schulhoff and Drake \cite{Sch15} assuming decay from the $^3S_1$ state. 

\section{Experimental method}
\label{Sec_ExpSetup}

The $^6{\rm He}$ atoms were produced via the $^7{\rm Li}(d,t)^6{\rm He}$ reaction. Up to 2$\times$10$^{10}$ $^6{\rm He}$ atoms per second were produced by bombarding a lithium target with an 18~MeV deuteron beam, delivered by the tandem Van de Graaff accelerator at the University of Washington \cite{Kne11}. 

The $^6{\rm He}$ atoms were pumped into a RF-discharge tube where a fraction ($\sim10^{-5}$) of the $^6$He atoms were brought to the $^3S_1$ metastable atomic state. The forward-going metastable atoms were then transversely-cooled, slowed by a Zeeman slower, and trapped in a magneto-optical trap,  all based on resonant excitation of the $^3S_{1}$ to $^3P_2$ electronic transition via 1083 nm laser light. Due to the small efficiency for pumping $^6{\rm He}$ atoms to the metastable state there was a considerable amount of ground-state $^{6}{\rm He}$ in the chamber hosting the MOT. To reduce events from non-trapped $^6{\rm He}$ atoms, the atoms in the MOT were periodically pushed by a laser beam into a measurement chamber through a 5~mm-diameter 30-mm-long aperture tube for differential pumping and recaptured by a second MOT. This $^{6}{\rm He}$ trap contained over 2100 atoms on average. 

During the measurement, the trapping lasers of this MOT were alternatively switched on and off with a $1:1$ duty cycle and a period of 100 $\mu$s. At the beginning of each off-cycle of 50~$\mu$s, the $\sim$50\% fraction of the atoms previously excited to the $^3P_{2}$ level quickly decay back to the $^3S_{1}$ level within the 100~ns lifetime of the excited state. Time correlation of the decay events with the switching cycle thus allowed us to isolate decays from $^{6}$He purely in the $^3S_{1}$ state (laser off) from decays with an $\sim$50\% admixture of the $^3P_{2}$ excited state (laser on). Meanwhile, the fast switching provided sufficient confinement of the atom cloud.

\begin{figure}[h!]
\centering
  \includegraphics[width=0.9\linewidth]{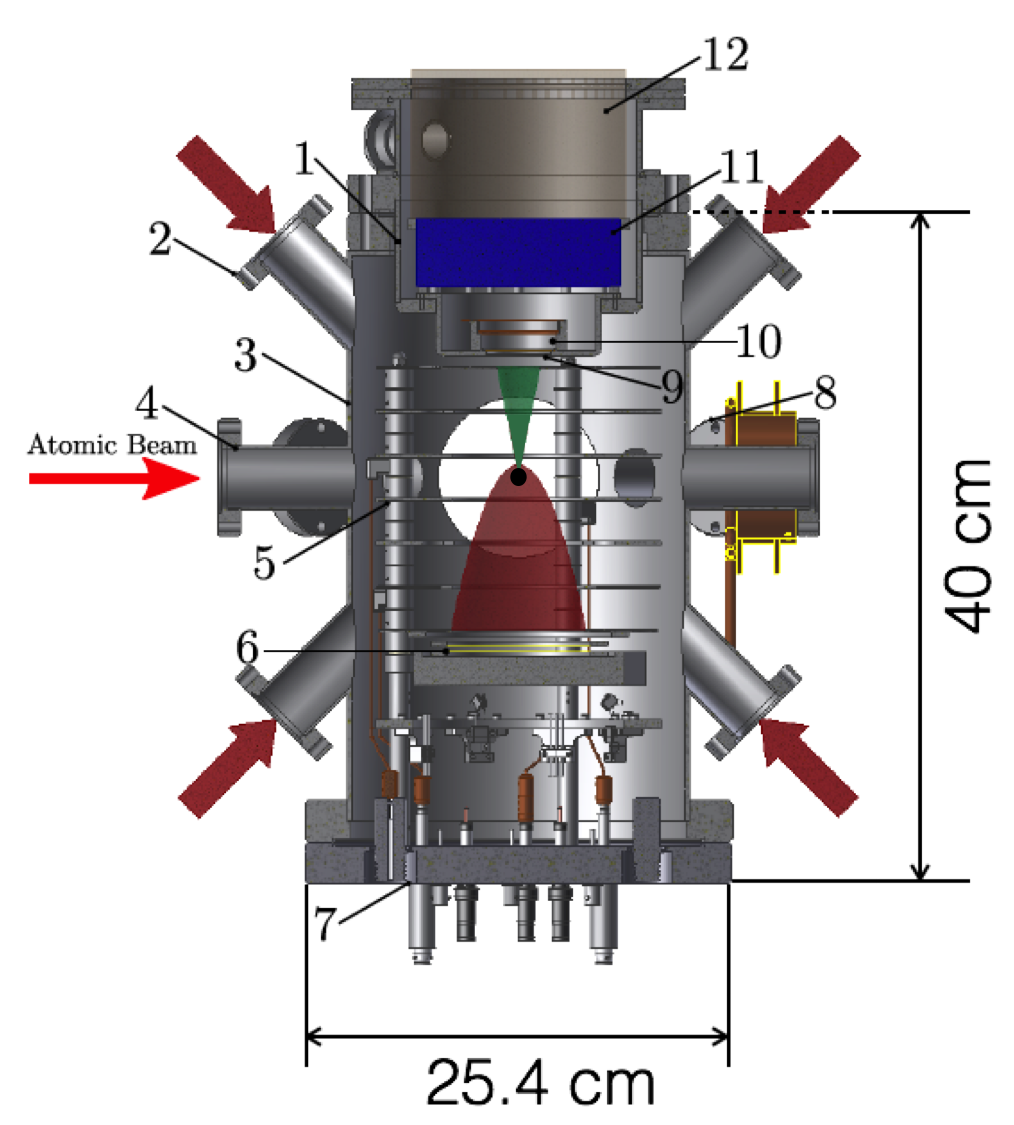}
\caption{Cross-section view of the detector system mounted on the measurement chamber. 1) re-entrant $\beta$-telescope housing, 2) trapping laser ports, 3) main chamber, 4) $^{6}$He transfer port, 5) electrode assembly, 6) micro-channel plate (MCP) recoil-ion detector, 7) 10~inch custom feedthrough flange for HV and MCP connections, 8) trap monitoring ports, 9) 127~$\mu$m Be foil, 10) multi-wire proportional
chamber (MWPC), 11) plastic scintillator, 12) lightguide to photo-multiplier tube.\label{Fig_DetectorSysSchematic}}
\end{figure}

We studied the charge-state distributions of the recoil $^{6}{\rm Li}$ ions by analyzing their time-of-flight (TOF) and energy spectra, which required detecting the $\beta$ particle and the recoil ion in coincidence. The configuration of the detection chamber is shown in Fig.~\ref{Fig_DetectorSysSchematic}. 

A $\beta$-telescope, consisting of a multi-wire proportional chamber (MWPC) and a scintillation detector, is placed above the trap. The scintillation detector measures the energy of the $\beta$ particle ($E_{\beta}$). The MWPC detects the entrance position of the particles and strongly suppresses $\gamma$-ray backgrounds triggering the scintillator when applying an appropriate coincidence gate between the two $\beta$ detectors. The MWPC runs with 1~atm ${\rm Ar-CO}_{2}$ ($9:1$ by volume) gas and is separated from the MOT vacuum by a $127~\mu {\rm m}$ thick 3.81-cm diameter beryllium window. 

A Micro Channel Plate (MCP) detector \cite{MCPpaper} is placed below the trap for detecting recoil ions and determining their hit positions with a resolution of 190~$\mu$m (FWHM). Electrodes are installed in-between the $\beta$-telescope and the MCP detector to create an electric field of $E \approx 1.3~{\rm kV/cm}$ and accelerate the recoil ions emitted from the trap towards the MCP detector. This enhances the ion-collection solid angle so that $\sim 85$\% of the $^{6}{\rm Li}^{+}$ and 100\% of the $^{6}{\rm Li}^{2+}$ and $^{6}{\rm Li}^{3+}$ ions are collected within the 75~mm diameter active area of the MCP. It also allows for the ions to have enough energy to trigger the MCP detector with a maximal detection efficiency ($\sim$50\%), independent of their initial charge state and recoil energy. 

\begin{figure}[h!]
  \centering
  \subfloat{\includegraphics[width=\linewidth]{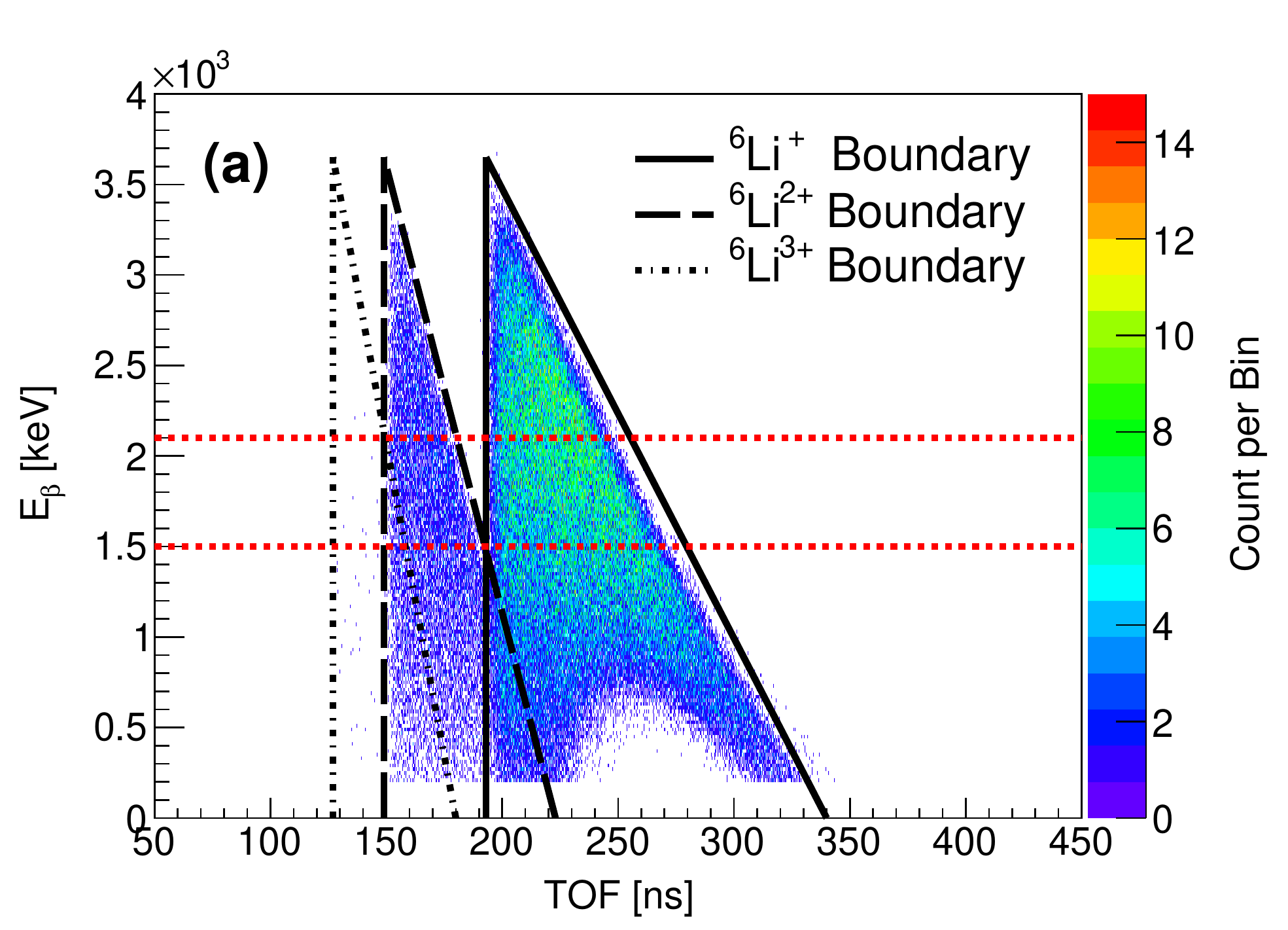}\label{Fig:E_beta_vs_TOFDat}}\\
  \subfloat{\includegraphics[width=\linewidth]{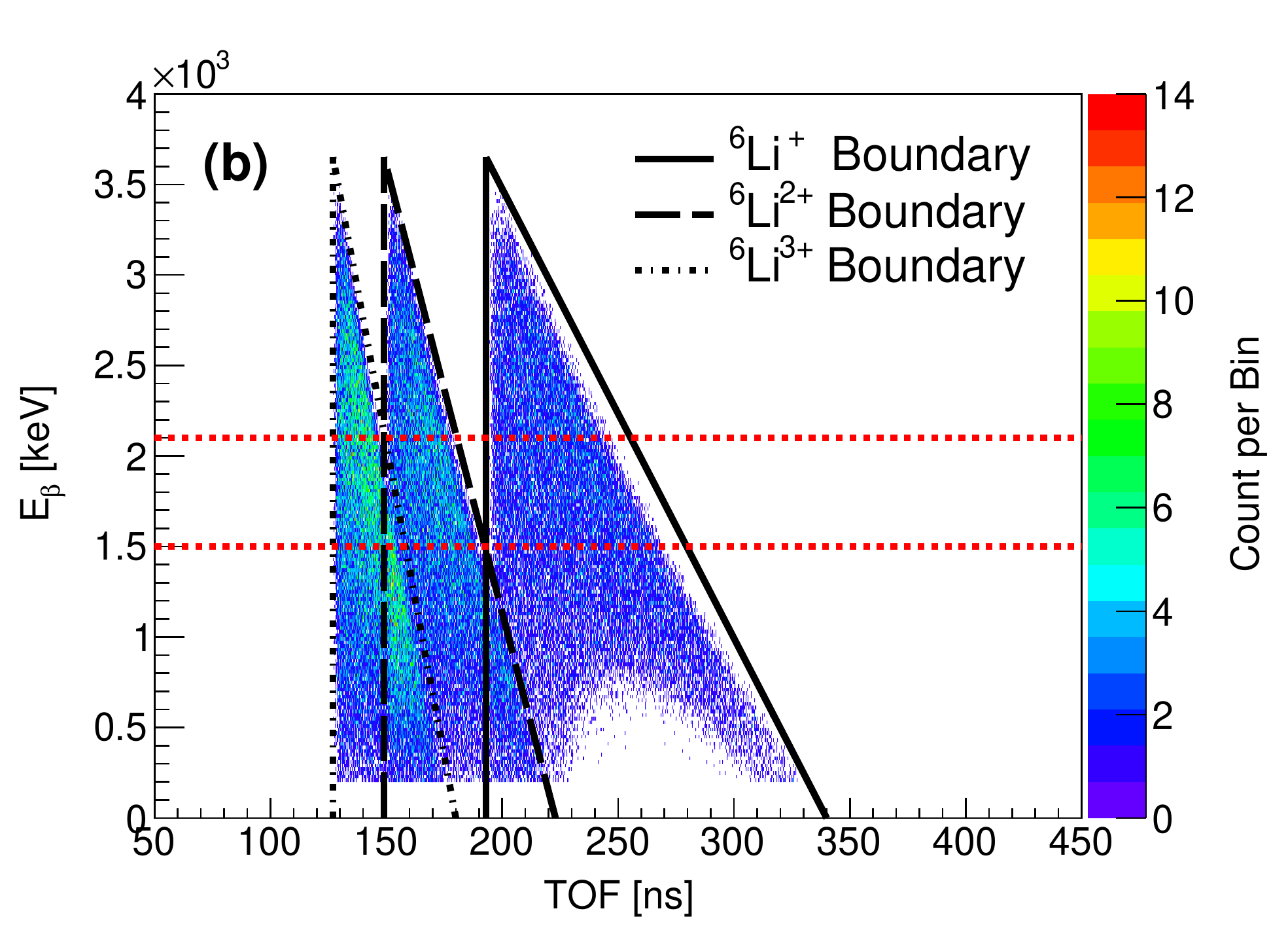}\label{Fig:E_beta_vs_TOFSim}}
\caption{$E_{\beta}$-vs.-TOF 2D histograms for (a) data from measurements and (b) a Monte Carlo simulation assuming equal charge-state fractions. The boundaries for the three charge states are drawn in black lines, and the minimal $E_{\beta}$ thresholds needed to separate the charge states are drawn in red lines. The lack of events near TOF=250~ns and $E_{\beta}<750$~keV is due to the fiducial cut on the MCP. \label{Fig:E_beta_vs_TOF}}
\end{figure}

The TOF measurement of the recoil ion is started by a scintillator signal and stopped by a MCP signal. The resolution of the TOF measurement is 820~ps (FWHM). $^{6}{\rm Li}$ ions in different charge states have different accelerations in the applied electric field, and are thus partially separated in TOF as shown in Fig.~\ref{Fig:E_beta_vs_TOF}. The overlaps between charge states can be avoided by applying a high enough threshold to the $\beta$-energy as indicated by the horizontal dashed lines. 

To fully reconstruct the initial energy of the recoil ion, the MOT position must also be determined. We ionize the trapped $^{6}$He atoms periodically (at $\sim$20~Hz) using a 2-ns pulsed beam of ultraviolet (337~nm) nitrogen laser that has sufficient photon energy to photo-ionize the $^3P_{2}$ state but not the $^3S_{1}$ state. 
The vertical coordinate (along the electric field direction) of the MOT is determined through the TOF of the photo-ions with respect to the laser pulse. 

A unique feature of the trapping of metastable ${\rm He}$ atoms is that it allows for monitoring the horizontal shape of the MOT via ``Penning-ions''. The latter are generated by collisions between neutral atoms in the non-perfect vacuum and metastable $^6{\rm He}$ atoms. 
The uncertainty of the start position of each $^{6}{\rm Li}$ recoil ion is limited in this data set by the size of the MOT which is 1.4~mm (FWHM). With the precisely measured MOT position, electric field strength, recoil-ion TOF and hit positions, the initial momenta of recoil ions are determined. The detector responses are calibrated daily, and the electric field and MOT position are monitored while the data are taken. The details of the construction, calibration and performance of the detector system are described in Ref.~\cite{Hong_Thesis}. 

Decays from non-trapped $^{6}$He atoms in the decay chamber and scattering of $\beta$ particles prior to their detection generate an undesired background. To suppress this background, we reconstructed the momentum of the emitted anti-neutrino using the measured $\beta$-momentum and ion-momentum, and evaluated the $Q$-value (the total kinetic energy released from the decay), which should be 3.5~MeV. Decays outside the MOT yield incorrect $Q$-values because the reconstruction assumes the events originate at the MOT. Our $Q$-value cut ($Q$-cut) accepted events with $3.072 \le Q \le3.858~{\rm MeV}$, so that $\sim 90$\% of the events from non-trapped decays were eliminated while the total data loss for events from trapped decays was less than 0.1\% \cite{Hong_Thesis}. The width of this cut was limited by the energy resolution in our scintillator.
In order to determine the contribution of the remaining events from non-trapped decays, we introduced large amounts of ground-state $^{6}{\rm He}$ atoms via a bypass pipe. 

\begin{figure}[h!]
  \centering
  \subfloat{\includegraphics[width=\linewidth]{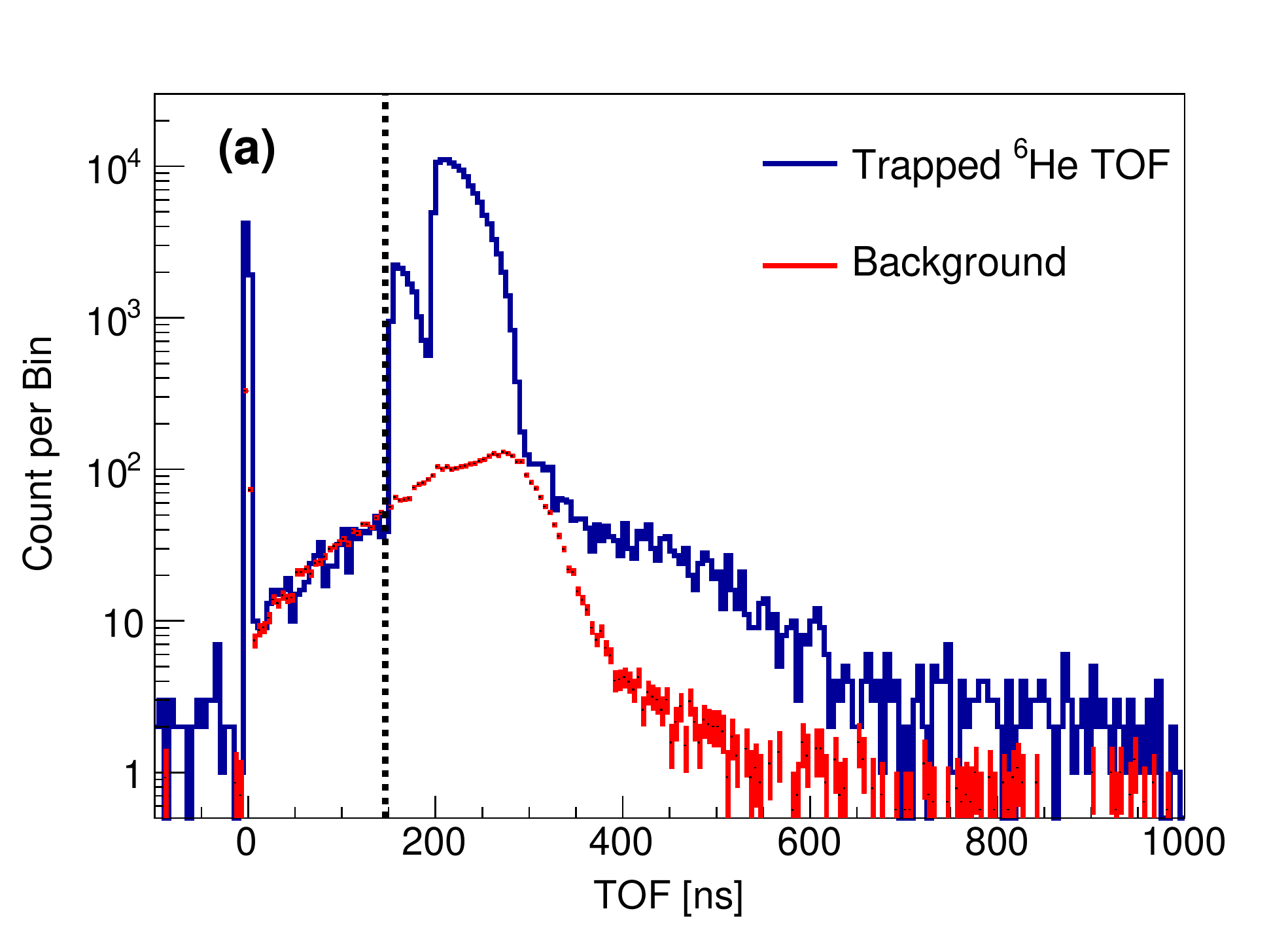}\label{Fig_ChargeStateTOFBkgSubQOff}}\\
  \subfloat{\includegraphics[width=\linewidth]{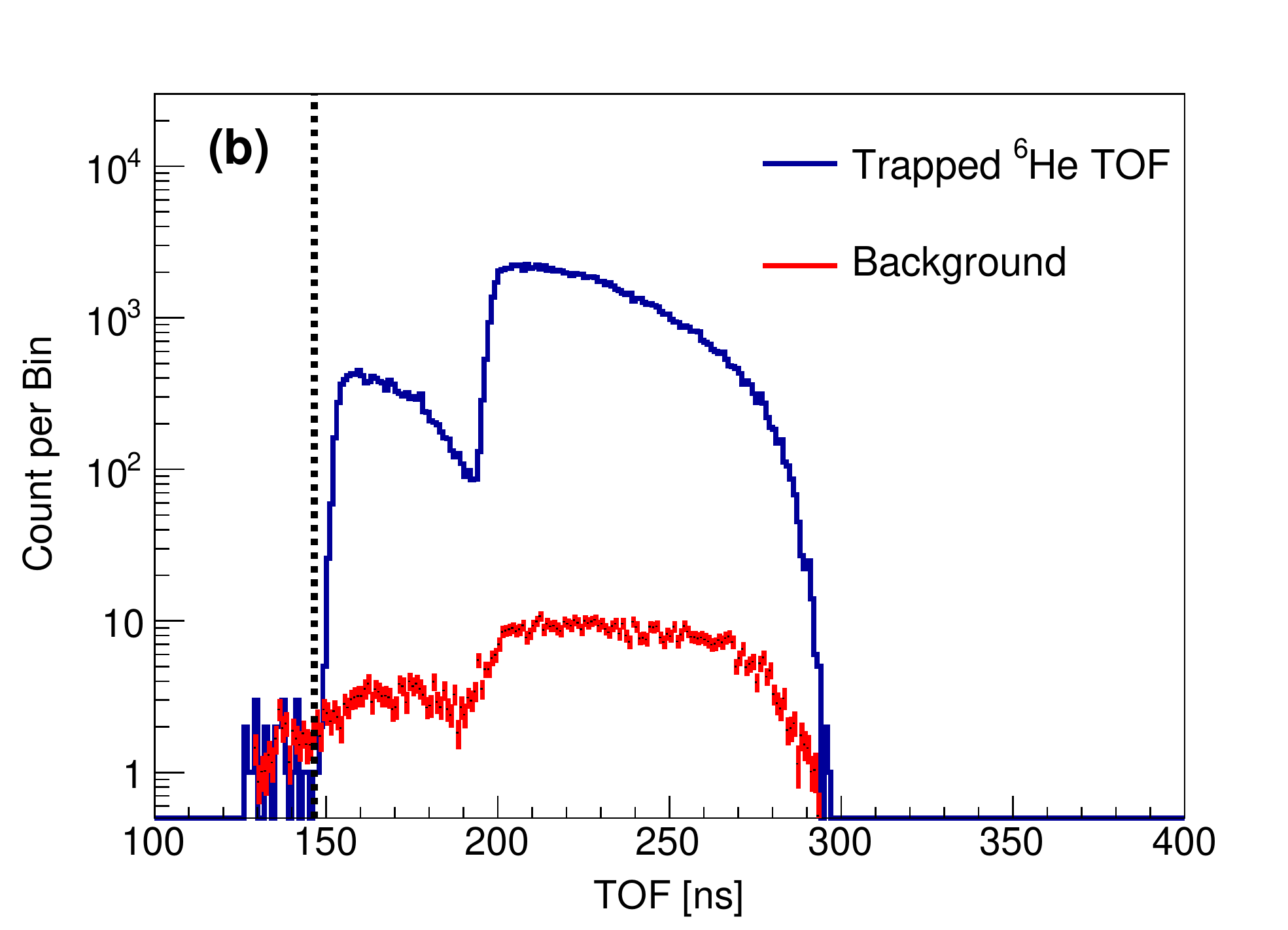}\label{Fig_ChargeStateTOFBkgSub}}
\caption{TOF spectra for laser-off data (a) without and (b) with the $Q$-cut. The normalized spectra from non-trapped atoms (red curves) for each case are overlaid. $E_{\beta}$ threshold was set to 1~MeV when generating these plots. The TOF region between 10 and 100 ns on (a) was used to normalize the curves. The sharp peak at time $\approx 0$ in spectra (a) corresponds to electrons which backscattered hitting both the MCP and $\beta$ detectors. The TOF leading edge for $^{6}{\rm Li}^{2+}$ is shown in black dashed lines in (a) and (b).}
\label{fig:bkg normalization}
\end{figure}

Fig.~\ref{fig:bkg normalization} shows a comparison of the TOF spectra. In Fig.~\ref{Fig_ChargeStateTOFBkgSubQOff} the data from the non-trapped $^{6}{\rm He}$ is normalized and overlaid with the data taken with the trap on. The shortest TOF for a $^{6}{\rm Li}$ ion emitted from the MOT is $T_{LE3}=127$~ns which is the TOF leading edge for $^{6}{\rm Li}^{3+}$. Events with TOF$<T_{LE3}$ must be from non-trapped $^{6}$He decays occurring closer to the MCP. Therefore, the normalization was chosen to match the TOF spectra without the $Q$-cut in the TOF region from 10~ns to 110~ns. The same normalization factor was applied to the TOF and ion-energy spectra after the $Q$-cut (shown in Fig.~\ref{Fig_ChargeStateTOFBkgSub}) and then these spectra were subtracted from those from the corresponding data run to remove the remaining events from untrapped $^6{\rm He}$ atoms. 

\section{Data Analysis and Results}
\label{sec:charge state}

As apparent from the data shown in Fig.~\ref{Fig:E_beta_vs_TOFDat} and Fig.~\ref{Fig_ChargeStateTOFBkgSub}, we clearly observe some shake-off fraction yielding $^{6}{\rm Li}^{2+}$ ions, however, we do not observe $^{6}{\rm Li}^{3+}$ ions above background and can therefore only set an upper limit. In the following, we will treat these two cases separately.

We first address the total fraction of $^{6}{\rm Li}^{3+}$ ions ($P_{3}$), regardless of their energies. $P_{3}$ is the ratio $N_{3}/N_\text{all}$ , where $N_{3}$ is the number of observed $^{6}{\rm Li}^{3+}$ ions, and $N_\text{all}$ is the number of observed $^{6}{\rm Li}$ ions in all charge states. It is important to choose an appropriate $E_{\beta}$ threshold to ensure a correct determination of $N_\text{all}$. As shown in Fig.~\ref{Fig:E_beta_vs_TOF}, some of the $^{6}{\rm Li}^{+}$ events with small $E_{\beta}$ are lost because they have too high transverse momentum to hit the active area of the MCP. Therefore, the $E_{\beta}$ threshold should be set high enough so that $^{6}{\rm Li}$ ions in all charge states above the $E_{\beta}$ threshold fall in the active area of the MCP. We chose 1~MeV as the $E_{\beta}$ threshold, and under this condition the event loss due to the finite size of the MCP fiducial area is then less than 0.1\% according to the Monte Carlo simulation. However, an $E_{\beta}$ threshold at 1~MeV is not high enough to separate $^{6}{\rm Li}^{3+}$ from $^{6}{\rm Li}^{2+}$ ions. Therefore, to obtain the correct $N_{3}$, we determined the number of events below the leading edge ($T_{LE2}=149~{\rm ns}$) of $^{6}{\rm Li}^{2+}$ TOF spectrum, and then corrected the $^{6}{\rm Li}^{3+}$ counts to include the counts overlapping with $^{6}{\rm Li}^{2+}$ events using a Monte Carlo simulation. Given the $E_{\beta}$ threshold at 1~MeV, the percentage of $^{6}{\rm Li}^{3+}$ ions that are below $T_{LE2}$ over all $^{6}{\rm Li}^{3+}$ ions is 85.5\%. (Alternatively, it is also possible to choose an $E_{\beta}$ threshold at 2.1~MeV so that the $^{6}{\rm Li}^{3+}$ ions are completely separated from the $^{6}{\rm Li}^{2+}$ ions, but such a high threshold rules out $\sim$75\% of the data, yielding poor statistics so it was not adopted.) 

The TOF spectrum with the $Q$-cut for the laser-off data is plotted in Fig.~\ref{Fig_ChargeStateTOFBkgSub}. The event counts in the $^{6}{\rm Li}^{3+}$ TOF region (TOF$<T_{LE2}$) are dominated by the background generated by the non-trapped $^{6}$He decays. A summary of results is shown in Table.~\ref{tab:Charge3Result}. The measurements are consistent with $P_{3}=0$ within 1 standard deviations, and the 90\% confidence levels are calculated only in the physical region where the count of events are greater than 0.

\begin{table}[!h]
\caption{Measured $^6{\rm Li}^{3+}$ ion charge fractions for $^6{\rm He}$ decays from atomic excited states, under the conditions that $E_{\beta}>1$~MeV. The negative $P_{3}$ values originate from the background subtraction. The 90\% confidence levels are calculated only in the physical region where $P_{3} > 0$.\label{tab:Charge3Result}}
\begin{ruledtabular}
\begin{tabular}{lccc}
Laser & $P_{3}$ $\times 10^{5}$ & $\Delta P_{3}$$\times 10^{5}$ & Upper limit $\times 10^{5}$\\
status & &  & 90\% C.L. \\
\hline
 On & $-1.2$ & 6.5 &  10 \\
 Off & $-0.6$ & 6.6 &  11 \\
\end{tabular}
\end{ruledtabular}
\end{table}

\begin{figure}[h!]
  \centering
  \includegraphics[width=\linewidth]{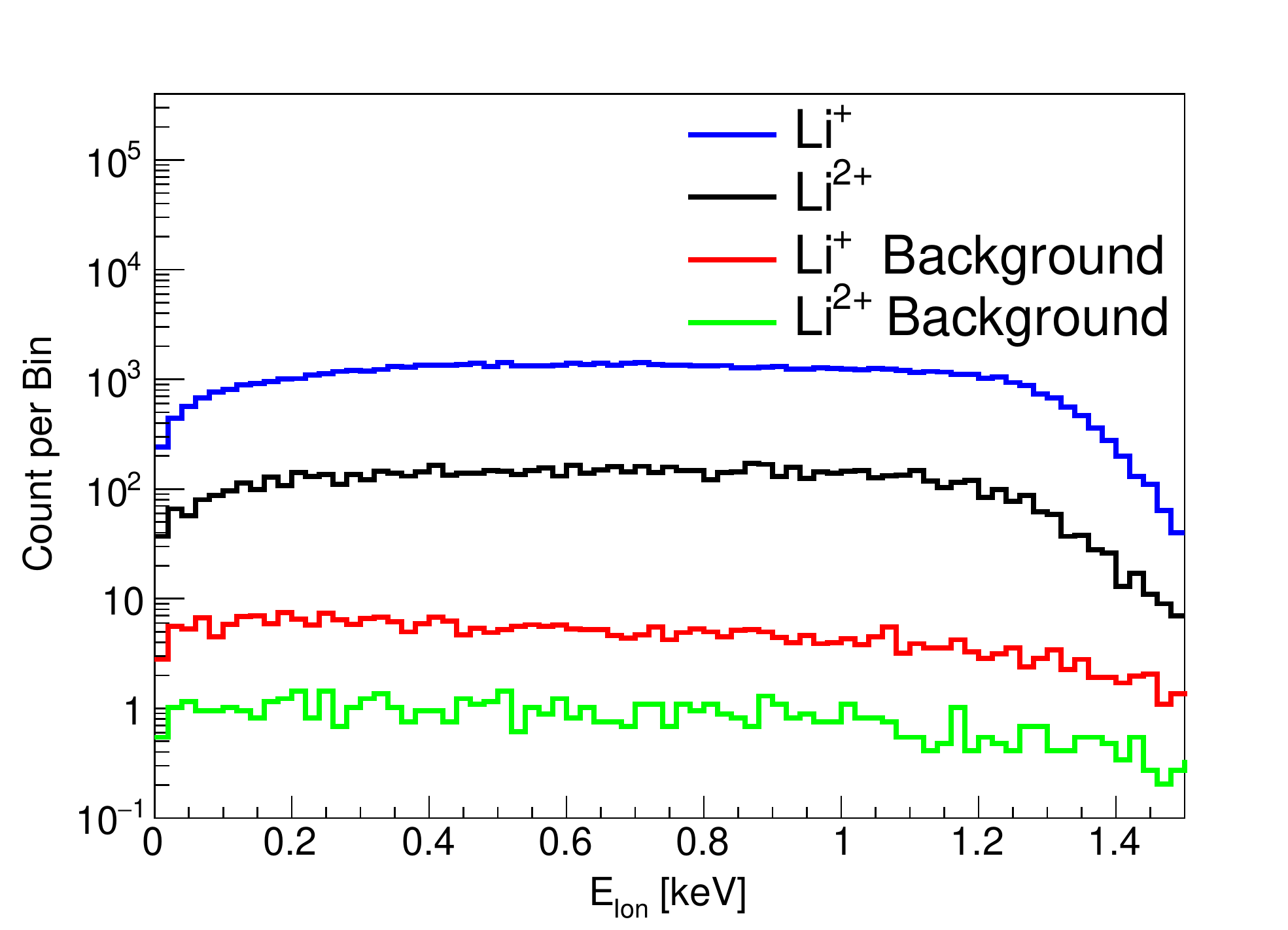}\label{fig:ChargeStateEIon1}
\caption{$^{6}{\rm Li}$ recoil ion initial energy distributions for Li$^{+}$ and Li$^{2+}$. Background spectra are also plotted.\label{fig:ChargeStateEIon}}
\end{figure}

Next, we studied the fraction of $^{6}{\rm Li}^{2+}$ ions and its dependence on the recoil-ion energy. In this study, the initial energy of the recoil ions ($E_{Ion}$) are reconstructed for each event, so it is necessary to determine the charge state of each ion with no ambiguity. Therefore, the minimal threshold on $E_{\beta}$, 1.5~MeV, was applied so that $^{6}{\rm Li}^{2+}$ ions are completely separated from $^{6}{\rm Li}^{+}$ in TOF as shown 
by the lower of the two horizontal red dashed lines in Fig.~\ref{Fig:E_beta_vs_TOF}. As discussed above, the upper limit for the probability of having a $^{6}{\rm Li}^{3+}$ ion is at the $10^{-4}$ level, approximately three orders of magnitude lower than the probability of having a $^{6}{\rm Li}^{2+}$ ion. Therefore, we neglected $^{6}{\rm Li}^{3+}$ ions in this study assuming that all events below the leading edge ($T_{LE1}=193$~ns) of $^{6}{\rm Li}^{+}$ TOF spectrum correspond to $^{6}{\rm Li}^{2+}$ ions. The $E_{Ion}$ spectra for $^{6}{\rm Li}^{+}$ and $^{6}{\rm Li}^{2+}$ with their corresponding normalized backgrounds are plotted in Fig.~\ref{fig:ChargeStateEIon}.

Theoretically, the probability distribution functions (PDFs) of $E_{Ion}$ in the two charge states are
\begin{align}
{\cal P}_{i}(E_{Ion})=\Phi(E_{Ion})\times P_{i}(E_{Ion})\times  \eta_{i}.
\label{eq:EIonSpectrum}
\end{align}
where $i=1,2$ indicates the charge state, and $P_{i}$ is the probability of having an ion in such charge state. $\Phi(E_{Ion})$ represents the part of the probability distribution function that is independent of charge state and depends only on the $\beta$-decay dynamics, {\em i.e.} the $\beta-\nu$ correlation coefficient $a_{\beta\nu}$. The detection efficiency, $\eta_{i}$, depends on the detection geometry, detector response functions and event reconstruction parameters and conditions. The expected energy dependence for $P_{i}(E_{Ion})$ is \cite{Sch15}:
\begin{align}
P_{i}=A_{i}+B_{i}\,E_{Ion},
\label{eq:ShakeOffProbability}
\end{align} 
where $A_{i}$ and $B_{i}$ are parameters. Taking the ratio of ${\cal P}_{2}$ and ${\cal P}_{1}$, one gets
\begin{align}
R(E_{Ion}) = \frac{(A_{2}+B_{2}E_{Ion})\times  \eta_{2}}{(A_{1}+B_{1}E_{Ion})\times \eta_{1}}.
\label{eq:ShakeOffProbRatio}
\end{align}
The $\Phi(E_{Ion})$ functions in the numerator and the denominator cancel each other and $R(E_{Ion})$ no longer depends on $a_{\beta\nu}$ explicitly. With the $E_{\beta}=1500$~keV threshold there is no event loss due to MCP fiducial area for both charge states $q=1$ and $q=2$. In principle, the ion detection efficiency $\eta_{i}$ depends on the charge state because $^{6}{\rm Li}^{+}$ and $^{6}{\rm Li}^{2+}$ have very different final energies ($\approx$13~keV versus $\approx$26~keV), and their position distributions on the MCP are also different. However, the difference in gain of the MCP for these two charge states is only 2.5\%. Due to a very low MCP charge threshold, it results in less than $2 \times 10^{-5}$ efficiency difference. Therefore, we made the assumption that $\eta_{2}=\eta_{1}$, so they cancel each other in Eq.~(\ref{eq:ShakeOffProbRatio}). (The systematic uncertainty generated by this approximation was studied using Monte Carlo simulations including the measured 0.82\% (RMS) variation of the MCP efficiency over the surface of the MCP.) Because $^{6}{\rm Li}^{3+}$ ions are neglected, the sum of $P_{1}(E_{Ion})$ and $P_{2}(E_{Ion})$ is 1, and thus Eq.~(\ref{eq:ShakeOffProbRatio}) becomes
\begin{align}
R(E_{Ion}) = \frac{A_{2}+B_{2}E_{Ion}}{1-A_{2}-B_{2}E_{Ion}}.
\label{Eq_ShakeOffFitFunc}
\end{align}

\begin{figure}[h!]
  \centering
  \includegraphics[width=\linewidth]{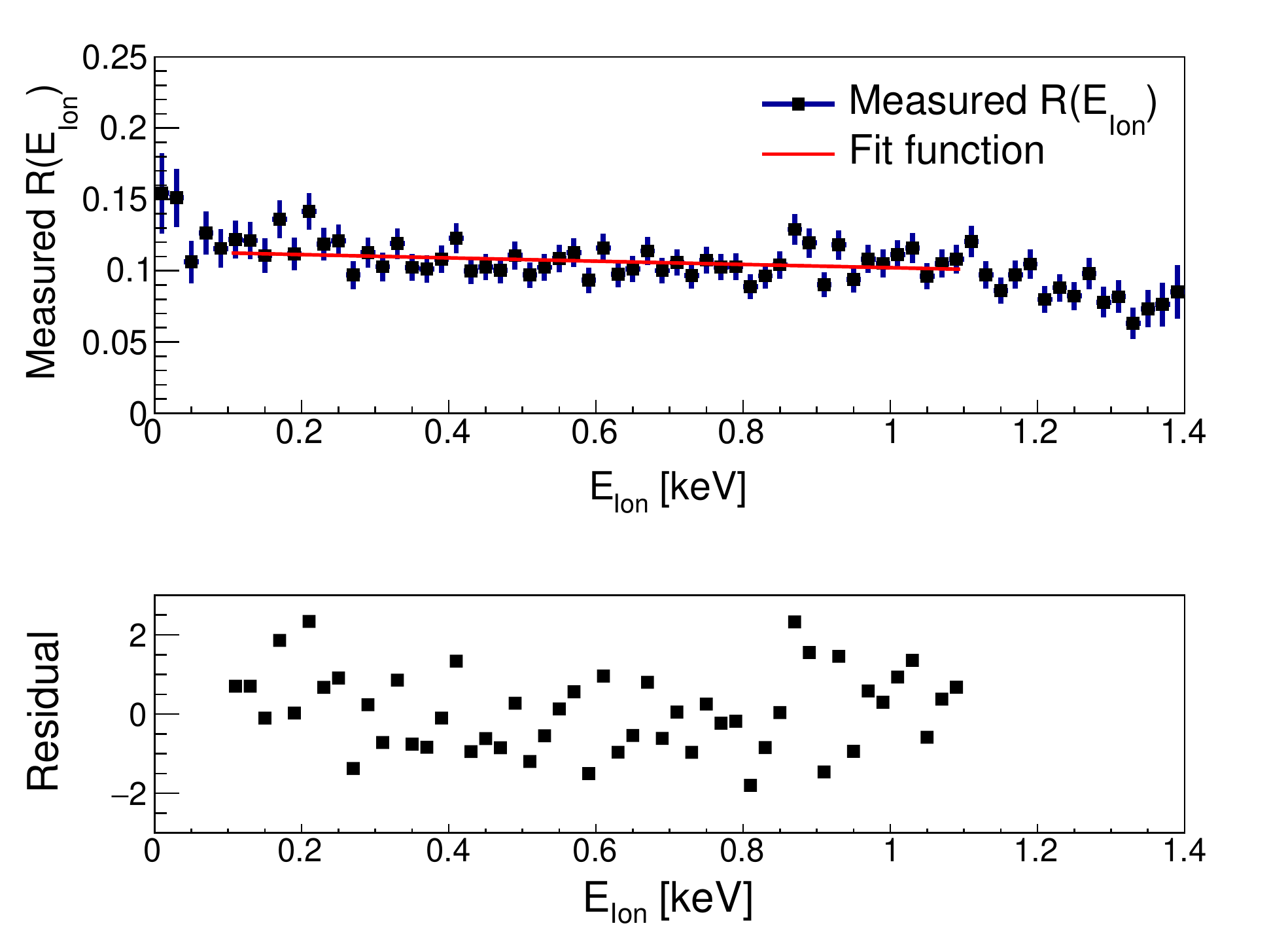}
\caption{Ratio between the $E_{Ion}$ spectra for the charge state 2 and 1, fit to Eq.~(\ref{Eq_ShakeOffFitFunc}) in the range 0.1~keV$<E_{Ion}<$1.1~keV for the laser-off data. Normalized residuals inside the fit region are plotted in the lower panel. \label{fig:ChargeStateEIonFit}}
\end{figure}

We take the ratio of the measured $E_{Ion}$ spectra (with background subtraction applied) for the charge states $q=2$ and $q=1$, and fit the ratio histogram to Eq.~(\ref{Eq_ShakeOffFitFunc}), as shown in Fig.~\ref{fig:ChargeStateEIonFit}. The fit region is chosen to be from 0.1~keV to 1.1~keV, in order to make the systematic uncertainties introduced by the assumptions made above negligible compared to the statistical uncertainty. $A_{2}$ and $B_{2}$ are fit parameters, and the fit results are listed in Table~\ref{tab:ChargeStateFitRes}. We also obtain the fractions of $^{6}{\rm Li}^{+}$ and $^{6}{\rm Li}^{2+}$ regardless of their energies by counting the total number of events above and below $T_{LE1}$. The results of the fractions for all three charge states of $^{6}{\rm Li}$ ions are summarized and compared to theoretical calculations \cite{Sch15} in Table~\ref{tab:total fractions II}. 
All the fractions determined in this experiment are obtained with an $E_{\beta}$ threshold. Applying an $E_{\beta}$ threshold, which is inevitable in this experiment, changes the energy spectrum of the recoil ions, and thus affects the fractions of $^{6}{\rm Li}$ ions in different charge states due to their dependences on the ion energies. Therefore, in Table~\ref{tab:total fractions II} the theoretical fractions are calculated based on the results of Ref.~\cite{Sch15} and applied with the same $E_{\beta}$ threshold as in our experiment.

\begin{table}[h!]
\caption{Fit results for $A_{2}$ and $B_{2}$ in Eq.~(\ref{Eq_ShakeOffFitFunc}).\label{tab:ChargeStateFitRes}}
\begin{ruledtabular}
\begin{tabular}{cccc}
Laser & $A_{2} \times 100$ &  $B_{2} \times 10^4 $ & $\chi^{2}$/dof\\
  &                       &  (${\rm keV}^{-1}$) &                       \\ 
\hline
 On & 10.1 $\pm$ 0.3 & -36 $\pm$ 42 & 45/48 \\
 Off & 10.2 $\pm$ 0.3 & -94 $\pm$ 42 &  49/48\\
\end{tabular}
\end{ruledtabular}
\end{table}

\begin{table}[!h]
\caption{Comparison of calculated versus measured $^6{\rm Li}$ ion charge fractions (in \%) for $^6{\rm He}$ decays from $^3S_1$ atomic metastable state.}
\begin{ruledtabular}
\begin{tabular}{lll}
Ion & \multicolumn{1}{c}{Theory\footnotemark[1] \cite{Sch15}} & \multicolumn{1}{c}{This work}\\
\hline
${\rm Li}^+$       & $88.63(2)$  & $90.5(1)$  \\
${\rm Li}^{2+}$  & $9.5(1)$      & $9.5(1)$ \\
${\rm Li}^{3+}$  & $1.9(1)$      & $\le 0.01$\\
\end{tabular}
\footnotetext[1]{When calculating the theoretical charge fractions for $^{6}{\rm Li}^{+}$ and $^{6}{\rm Li}^{2+}$, the averaged ion energy is determined using a 1.5~MeV $\beta$-energy threshold as used in the experiment. In this case, $\langle E_{ion}\rangle=0.74$~keV. For $^{6}{\rm Li}^{3+}$, $\beta$-energy threshold is 1~MeV, and $\langle E_{ion}\rangle=0.74$~keV as well.}
\label{tab:total fractions II}
\end{ruledtabular}
\end{table}

In order to understand the systematic shifts of $A_{2}$ and $B_{2}$ caused by approximations like assuming $\eta_{1}$ and $\eta_{2}$ to be identical, we ran Monte Carlo simulations with the values of $A_{2}$ and $B_{2}$ from Ref.~\cite{Sch15}. In the simulations, we modeled the experimental parameters, such as the detector geometry and response functions, and the efficiencies and electric field, as close as possible to the experimental setup. The simulated data were processed in the same way as that used for the experimental data. There is no significant deviation of the extracted $B_{2}$ value from its input value, while the extracted $A_{2}$ deviates from its input by $-7(4)\times10^{-4}$. Implementation of the 0.82\% spatial variation of the MCP efficiency does not result in significant deviations of $A_{2}$ and $B_{2}$. We also studied how the fit values of $A_{2}$ and $B_{2}$ change with respect to parameters used in the ion-energy reconstruction, and the systematic uncertainties associated with these parameters. The corresponding non-negligible systematic uncertainties of $A_{2}$ and $B_{2}$ are listed in Table~\ref{tab:SysUncertainty}, and in total are smaller than the statistical uncertainties listed in Table~\ref{tab:ChargeStateFitRes}. Note that these systematic uncertainties are all related to the ion-energy calculation, so they do not affect the determination of the energy-integrated charge-state fraction listed in Table~\ref{tab:total fractions II}. 

\begin{table}[h!]
\caption{Systematic shifts and uncertainties of $A_{2}$ and $B_{2}$.\label{tab:SysUncertainty}}
\begin{ruledtabular}
\begin{tabular}{cccc}
 & Shift $A_{2} \times 100$ &  $\Delta A_{2} \times 100$  &   $\Delta B_{2} \times 10^{4}$ \\
  &                                       &                                              &       (${\rm keV}^{-1}$)                \\ 
\hline
 Approximations\footnotemark[1]  &  $-0.07$ & 0.04  & $<$6\\
 Vertical MOT Pos. & & 0.066  & 6.1     \\
 Electric field  &       & 0.018  & 3.1     \\
 TOF Origin    &      & 0.055  & 4.6     \\
 \hline
 Total & $-0.07$ & 0.1 & 10\\
\end{tabular}
\footnotetext[1]{Uniform electric field, uniform efficiency of the MCP over its surface, point-like MOT.}
\end{ruledtabular}
\end{table}


\section{Discussion}
\label{sec:discussion}

A comparison of our measurements to calculations for the decay of $^6{\rm He}$ from its atomic metastable state is shown in Table~\ref{tab:total fractions II}. The $^{6}{\rm Li}^{3+}$ and $^{6}{\rm Li}^{+}$ fractions measured in this experiment have small ($\sim 2$\%) but significant ($\sim19 \sigma$ for $^{6}{\rm Li}^{3+}$ and $^{6}{\rm Li}^{+}$) discrepancies with the theoretical calculations of Ref.~\cite{Sch15}. The calculation for the $^{6}{\rm He}$ decays from the atomic ground state similarly over-predicts the $^{6}{\rm Li}^{3+}$ fraction measured by Carlson {\em et al.}~\cite{Car63} as shown in Table~\ref{tab:total fractions}. It is possible that there is a missing consideration in the calculation that systematically leads to a higher $^{6}{\rm Li}^{3+}$ fraction for both initial atomic states. In Ref.~\cite{Sch15} the charge state of the $^{6}{\rm Li}$ ion is determined solely by the final energy of the two orbital electrons, $E_{\rm tot}=E_1+E_2$, with respect to the ionization energies for 1 and 2 electrons, $E_{\rm ion-1}$ and $E_{\rm ion-2}$: 
\begin{eqnarray}
{\rm if}&&E_{\rm tot} \le E_{\rm ion-1} \rightarrow {^{6}{\rm Li}^{+}}\nonumber \\
{\rm if}&E_{\rm ion-1} \le &E_{\rm tot} \le E_{\rm ion-2} \rightarrow {^{6}{\rm Li}^{2+}} \nonumber \\
{\rm if}&&E_{\rm tot} > E_{\rm ion-2} \rightarrow {^{6}{\rm Li}^{3+}}. \nonumber 
\end{eqnarray}
In considering possible sources for the discrepancy we note, for example, that the condition $E_{\rm tot} > E_{\rm ion-2}$ could be met without double ionization if one of the electrons takes away a significant fraction of the energy as kinetic energy, leaving the other electron bound. This could lead to the systematic overestimation of the $^{6}{\rm Li}^{3+}$ probability as observed.

In contrast to the integrated charge-state fractions, the ion-energy dependencies of the $^{6}{\rm Li}$ ion charge-state fractions (the $B$ parameters) are of concern for the determinations of the $\beta-\nu$ correlation coefficient. The experiment of Ref.~\cite{Car63} was performed with the same method and apparatus used in Ref.~\cite{Joh63} to determine the $\beta-\nu$ correlation coefficient. As shown in Table~\ref{tab:comparison}, there is a significant discrepancy between the $B$ parameters measured by Carlson {\em et al.} and the ones calculated in Ref.~\cite{Sch15}. While we find a plausible explanation for some of the differences between theory and experiment for the overall Li-ion fractions, as stated above, it is more difficult to understand how the $B$ factors could be a factor of $\sim 7$ smaller in the calculations. The $B$ values measured by Carlson {\em et al.} are close to a naive prediction ignoring a cancellation that takes place between $nS$ and $nP$ final state configurations shown in detail in Ref.~\cite{Sch15}. Johnson et al. were aware of the cancellation and the fact that their measured $B$ factors disagreed with the more accurate calculation~\cite{Joh63}, but at the time they reported concerns about the difficulty of including a large enough set of states in their calculation. There is an implicit suggestion that the discrepancy should not be taken seriously because the calculation was incomplete. No such concerns exist for the recent calculation of Schulhoff and Drake~\cite{Sch15}, so we conclude that this suggests either an unaccounted-for experimental issue or a failure of the framework for the calculation. Changing the $B$ parameters from their measured value to zero affects the determination of $a_{\beta\nu}$ only by $0.6$\% which is smaller than the $1$\% uncertainty claimed by Ref.~\cite{Joh63}. In the present context of trying to achieve more precise determinations, the issue is more important.

\begin{table}[h!]
\caption{Comparison of calculations \cite{Sch15} to previous measurements \cite{Car63} of charge-distribution probabilities from the electronic ground state of $^{6}{\rm He}$.}
\begin{ruledtabular}
\begin{tabular}{lllll}
& \multicolumn{2}{c}{Theory \cite{Sch15}} & \multicolumn{2}{c}{Experiment \cite{Car63}}\\
Ion  & $A\times 100$ & $B \times 10^4$       & $A\times 100$ &$B \times 10^4$ \\ 
      &                       &  (${\rm keV}^{-1}$) &                       & (${\rm keV}^{-1}$)\\ 
\hline
${\rm Li}^+$  & $89.03(2)$  & $-6.17(2)$ & $89.9(2)$ & $-45(7)$ \\
${\rm Li}^{2+}$  & $9.7(1)$ & $+5.8(1)$ & $10.1(2)$ & $+42(7)$ \\
${\rm Li}^{3+}$  & $1.2(1)$ & $+0.34(14)$ & $0.018(15)$ & $+0.33(13)$ \\
\end{tabular}
\label{tab:comparison}
\end{ruledtabular}
\end{table}

Our results for the  $A$ and $B$ parameters for the $^{6}$He decays from the atomic metastable state and the corresponding theoretical calculations are shown in Table~\ref{tab:comparison-metastable}. Unfortunately, our results don't have the statistical power to claim a precision test of the calculations, in particular for the $B$ parameter. We also calculated the parameters $A_{2}$ and $B_{2}$ for decays from initial atomic state $^3P_{2}$, based on the $A_{2}$ and $B_{2}$ values in Table~\ref{tab:ChargeStateFitRes} for the laser-on case (mixture of 50\% $^3S_{1}$ and 50\% $^3P_{2}$) and laser-off case (pure $^3S_{1}$). The results together with the final results for the $^3S_{1}$ initial states are summarized in Table~\ref{tab:ChargeStateFitResFinal}. The results for these two initial states are not significantly different from each other. 

\begin{table}[h!]
\caption{Same as Table~\ref{tab:comparison} for the decay of $^{6}{\rm He}$ from its metastable state from this work.}
\begin{ruledtabular}
\begin{tabular}{lllll}
& \multicolumn{2}{c}{Theory \cite{Sch15}} & \multicolumn{2}{c}{This work \footnotemark[1]}\\
Ion  & $A\times 100$ & $B \times 10^4$       & $A\times 100$ &$B \times 10^4$ \\ 
      &                       &  (${\rm keV}^{-1}$) &                       & (${\rm keV}^{-1}$)\\ 
\hline
${\rm Li}^+$  & $88.711(3)$  & $-11.06(0)$ & $89.9(3)(1)$ & $94(42)(10)$ \\
${\rm Li}^{2+}$  & $9.42(7)$ & $+10.39(7)$ & $10.1(3)(1)$ & $-94(42)(10)$ \\
${\rm Li}^{3+}$  & $1.86(7)$ & $+0.74(148)$ & - & - \\
\end{tabular}
\footnotetext[1]{Systematic shifts for $A_{2}$ is included. The number in the first parenthesis is the statistical uncertainty, and the number in the second parenthesis is the systematic uncertainty. }
\label{tab:comparison-metastable}
\end{ruledtabular}
\end{table}

\begin{table}[h!]
\caption{Results for parameters $A_{2}$ and $B_{2}$, for $^3S_{1}$ and $^3P_{2}$ initial states.\label{tab:ChargeStateFitResFinal}}
\begin{ruledtabular}
\begin{tabular}{ccc}
Initial State & $A_{2} \times 100$ &  $B_{2} \times 10^4 $  \\
 &                       &  (${\rm keV}^{-1}$)                \\ 
\hline
 $^3S_{1}$ & 10.1(3)(1) & -94(42)(10)\\
 $^3P_{2}$ & 10.0(7)(1) & 22(94)(10)  \\
\end{tabular}
\footnotetext[1]{The number in the first parenthesis is the statistical uncertainty, and the number in the second parenthesis is the systematic uncertainty. }
\end{ruledtabular}
\end{table}

Additionally, in measurements of the $\beta-\nu$ correlation, where the TOF spectrum is fitted to templates generated by Monte Carlo simulations and the charge state groups are not completely separated in TOF, both the $A$ and $B$ parameters can affect the fit result. We have used a Monte Carlo simulations to study how the uncertainties of $A_{2}$ and $B_{2}$ translate into the uncertainty of the $\beta-\nu$ correlation coefficient. Based on the experimental uncertainties listed in Table~\ref{tab:comparison-metastable}, $A_{2}$ results in a 0.3\% relative uncertainty of $a_{\beta\nu}$, and $B_{2}$ results in a 0.6\% relative uncertainty. Therefore, the results from this paper are sufficient for an experiment aiming at determining $a_{\beta\nu}$ to 1\%. Because the charge-state analysis and the $a_{\beta\nu}$ analysis can use the same data set, the uncertainties of $A_{2}$ and $B_{2}$ will be improved as more data are taken to achieve better than 1\% uncertainties on $a_{\beta\nu}$. 

\section{Conclusions}

We have measured the $^{6}{\rm Li}$ ion charge-state fractions for $^{6}{\rm He}$ decays from atomic metastable state $^3S_1$. 
The overall fractions for $^{6}{\rm Li}^{+}$  and $^{6}{\rm Li}^{3+}$ (Table~\ref{tab:total fractions II}) show small but significant disagreement with the recent theoretical calculation of Ref.~\cite{Sch15}. We discuss a plausible explanation. 

We also point out that there is no satisfactory explanation for a large discrepancy between the same calculation and the results of Carlson et al.\cite{Car63} for the Li-ion energy dependence of the fractions from decays of the electronic ground state, suggesting either an unaccounted-for experimental issue or a failure of the framework for the calculation.

The $A$ and $B$ parameters in the ion-energy dependent charge-state fraction expression (Eq.~(\ref{eq:ShakeOffProbRatio})) were also determined (Table.~\ref{tab:ChargeStateFitResFinal}), and the precision is sufficient for a determination of $a_{\beta\nu}$ with 1\% relative precision. The precisions of the $A$ and $B$ parameters can be improved as more data for the $\beta-\nu$ correlation measurement are taken. 

\label{sec:conclusion}

\begin{acknowledgments}

This work is supported by the Department of Energy, Office of Nuclear Physics, under contract numbers DE-AC02-06CH11357
and DE-FG02-97ER41020. This work is also supported in part by the U.S. National Science Foundation under Grant No. PHY-11-02511. We thank Gordon Drake for many useful discussions.

\end{acknowledgments}
\raggedright
\bibliography{thebibliography}
\end{document}